\documentclass[aps,prb,preprint,showpacs]{revtex4}
\usepackage{graphicx}
\begin{document}

\title{Magnetotransport  in polycrystalline La$_{2/3}$Sr$_{1/3}$MnO$_{3}$ thin films of controlled granularity}

\author{P. K. Muduli, Gyanendra Singh, R. Sharma, R. C. Budhani}

\email{rcb@iitk.ac.in}
\affiliation{Condensed Matter - Low Dimensional Systems
Laboratory, Department of Physics, Indian Institute of Technology
Kanpur, Kanpur - 208016, India}

\date{\today}
\begin{abstract}

\baselineskip=1.5cm
 Polycrystalline La$_{2/3}$Sr$_{1/3}$MnO$_{3}$
(LSMO) thin films were synthesized by pulsed laser ablation on
single crystal (100) yttria-stabilized zirconia (YSZ) substrates
to investigate the mechanism of magneto-transport in a granular
manganite. Different degrees of granularity is achieved by using
the deposition temperature (T$_{D}$) of 700 and 800 $^{0}$C.
Although no significant change in magnetic order temperature
(T$_C$) and saturation magnetization is seen for these two types
of films, the temperature and magnetic field dependence of their
resistivity ($\rho$(T, H)) is strikingly dissimilar. While the
$\rho$(T,H) of the 800 $^{0}$C film is comparable to that of
epitaxial samples, the lower growth temperature leads to a
material which undergoes insulator-to-metal transition at a
temperature ( T$_{P}$ $\approx$ 170 K) much lower than T$_C$. At T
$\ll$ T$_P$, the resistivity is characterized by a minimum
followed by ln $\emph{T}$ divergence  at still lower temperatures.
The high negative magnetoresistance ($\approx$ 20$\%$) and ln
$\emph{T}$ dependence below the minimum are explained on the basis
of Kondo-type scattering from blocked Mn-spins in the
intergranular material. Further, a striking feature of the T$_D$ =
700 $^{0}$C film is its two orders of magnitude larger anisotropic
magnetoresistance (AMR) as compared to the AMR of epitaxial films.
We attribute it to unquenching of the orbital angular momentum of
3d electrons of Mn ions in the intergranular region where crystal
field is poorly defined.

\end{abstract}
\pacs{XX-XX, XX-XX-XX}
\maketitle
\clearpage \baselineskip=1.5cm
\section{Introduction}

Low field magnetoresistance (LFMR) in half-metallic granular
systems has attracted  considerable scientific interest because of
the physics of spin-polarized tunneling  which encompasses several
energy scales, as well as its potential technological
applications. Polycrystalline thin films of magnetite \cite{gong},
chromium dioxide \cite{coey},
manganites\cite{hwang,shreekala,gupta,li} and double perovskites
\cite{kobayashi,sarma,serrate} are  the most studied systems of
this class. It has also been shown that the magnetoresistive
properties of granular ferromagnetic metals can change
significantly due to spin-dependent scattering at the grain
boundries\cite{milner}. However, the grain boundary
magnetoresistance in manganites may be different from the
magnetoresistance (MR) of granular band ferromagnets such as Fe,
Co and Ni. In the former case, the magnetic structure at the
surface of grains is likely to be chaotic with blocked Mn-spins.
This is because of the propensity of surfaces to loose oxygen
which can render the double exchange interaction, the source of
magnetism and metallic conduction in these systems, weaker at the
surfaces and interfaces. Therefore, beside the usual tunneling
through insulating barriers, second order tunneling encompassing
interfacial spins at the grain boundaries may be important in
setting the MR of granular manganites\cite{Lee,hwang}.
Furthermore, when the granules are small, the Coulomb charging
energy may also be a relevant energy scale at low-temperatures.
The intergranular Coulomb blockade scenario has been invoked in
some cases to understand the low-temperature upturn in the
resistivity of granular manganites\cite{balcels, garcia,ju}. The
validity of such a hypothesis, however, needs to be examined by
looking for voltage threshold for conduction and non-linear
transport, which are important hallmarks of Coulomb blockade.

 Another important issue concerning the MR in granular
manganites is the anisotropic magnetoresistance (AMR). The AMR in
clean double exchange manganites has many contrasting features
compared to that in transition metal ferromagnets and alloys where
it is firmly believed to be due to spin-orbit interaction. In most
conventional metallic ferromagnets, the resistivity $\rho _ \bot $
is found to be smaller than $\rho _\parallel  $  and AMR decreases
on approaching the T$_C$, where $\rho _\parallel  $ and $\rho _
\bot $ denotes the resistivity for magnetic field parallel and
perpendicular to the current respectively. In contrast, manganites
typically display the inequality $\rho _\parallel   < \rho _ \bot
$. Also, the AMR is weakly temperature dependent, except for a
peak near the Curie temperature T$_C$\cite{ziese,yau,donnell}.
While a common mechanism may be responsible for the high-field AMR
in single crystal and polycrystalline manganites, at low fields
AMR of the latter may contain extrinsic contribution originating
from grain boundary magnetization anisotropy. Furthermore, since
the peak in temperature dependence of AMR of manganites has been
explained in terms of the unquenching of orbital moments due to
the enhanced Jahn-Teller (JT)  distortions near
T$_C$\cite{infante}, an enhanced AMR effect is expected in
polycrystalline samples where double exchange is suppressed at the
grain boundaries making JT electron-lattice coupling  more
relevant.

Here we report a detail study of both low-field magnetoresistance
and  anisotropic magnetoresistance  in  thin films of
La$_{2/3}$Sr$_{1/3}$MnO$_{3}$ (LSMO), the most promising manganite
for room temperature applications. These films were deposited
using an excimer laser based pulsed laser ablation technique  on
single crystal (100) yttria-stabilized zirconia (YSZ) and (100)
SrTiO$_3$ (STO) substrates. Also different deposition temperatures
(T$_{D}$ = 700 and 800 $^{0}$C) were used to control the granular
texture of the films. The crystalline quality and surface
morphology were well characterized through X-ray diffraction and
atomic force microscopy (AFM) techniques. The effect of
granularity on magnetic nature of the films was also analyzed by
observing the temperature and field dependence of magnetization. A
phenomenological model was introduced to explain the temperature
dependent resistivity in case of polycrystalline films deposited
at the lower temperature. The films were presumed to be a
two-component medium with ferromagnetic metallic intragrain and
spin-glass-type dirty intergrain metal. The low-temperature
up-turn in the resistivity was found to scale as $\rho$ $\sim$ ln
$\emph{T}$ representing a Kondo-type behavior in the dirty metal.
The large anisotropic magnetoresistance (AMR) were explained
through spin-orbit coupling and grain boundary effects.

\section{Experimental procedure}
Thin films of LSMO were fabricated by Pulsed Laser Deposition
(PLD) method as described in detail elsewhere \cite{senapati}. A
KrF excimer laser (wavelength $\lambda$ = 248 nm and  pulse width
$\approx$ 20 nanosecond) was used for ablation. Films of typically
100 nm in thickness were grown on (100) SrTiO$_{3}$ (STO) and
(100) yttria-stabilized zirconia (YSZ) substrates in the
temperature (T$_{D}$) range of 700 to 800 $^{0}$C in 0.4 mbar
oxygen pressure. A growth rate of $\approx$ 1.3 ${\AA}$/sec was
realized by firing the laser at 5 Hz with fluency of $\sim$ 3
J/cm$^{2}$/pulse on the target surface. After deposition, the
films were cooled down to room temperature at the rate of 10
$^{0}$C/min under an atmospheric oxygen pressure. The surface
morphology of the films was examined using an atomic force
microscope (AFM) where as for their  crystallographic structure we
have used a  X-ray diffractometer (PANalytical X'Pert PRO)
equipped with a Cu K$_{\alpha1}$ source. For electrical
measurements, samples were patterned in a four-probe geometry with
effective area of  200 $\mu $m $\times$ 1000 $\mu$m using optical
lithography and Ar$^{+}$ ion milling.  The resistivity and
magnetoresistance measurements were performed in a constant
current mode using a precision programmable DC current source
(Keithley 224), a temperature controller (Lakeshore 311) and a
nanovoltmeter (HP 34420 A nanovolt/micro-ohm meter). For the
measurements of magnetization, we have used a superconducting
quantum interference device (SQUID) based magnetometer (Quantum
Design MPMS-XL5).

\section{Results and discussion}
\subsection{Microstructural properties}

We first quantify the extent of granularity in our films using
X-ray diffraction and AFM techniques. Polycrystalline nature of
the LSMO films grown on YSZ at 700 and 800 $^{0}$C becomes clear
from the diffraction profiles  shown in  Fig. 1(a) and (b)
respectively. Although the lattice parameter  of YSZ (a$_{YSZ}$ =
0.514 nm) is very large compared to that of LSMO (a$_{LSMO}$ =
0.388 nm), the face diagonal of LSMO cubic cell is only 6.7 $\%$
shorter than a$_{YSZ}$, which should facilitate  a preferential
(110) oriented growth on (100) YSZ substrate. However, the
$\theta$ - 2$\theta$ scan shown in Fig. 1 reveals the presence of
(00l), (0kl) and (hkl) reflections in addition to (hk0),
confirming a truly polycrystalline growth. We also note that the
X-ray intensity ratio I(002)/I(110) for film deposited at 700  and
800 $^{0}$C is $\approx$ 0.14 and $\approx$ 2.99 respectively. The
smaller value of the ratio in case of films deposited at the lower
temperature imply more random texture as compared to that in the
film deposited at 800 $^{0}$C \cite{choi}.

The surface morphology of films deposited at 700  and 800 $^{0}$C
is shown in Fig. 2(a) and (b) respectively. The  AFM image of the
film with T$_{D}$ = 800 $^{0}$C shows granular morphology with
spherical grains of average diameter  $\approx$ 50 nm. The film
with T$_{D}$ = 700 $^{0}$C is seen to  contain ellipsoidal grains
with average length $\approx$ 80 nm and width $\approx$ 30 nm. The
rms roughness of the films was found out to be $\approx$ 8.59 nm
and $\approx$ 7.85 nm over an area of  5 $\times$ 5 $\mu$m$^{2}$
for film deposited at 700  and 800 $^{0}$C respectively.

\subsection{Electrical resistivity and its temperature and magnetic field dependence}
Epitaxial films of LSMO are characterized by metallic conduction
with a low ($\approx$ 110 $\mu\Omega$-cm) residual
resistivity\cite{senapati}. Interestingly, unlike the other
double-exchange manganites the conduction in LSMO remains metallic
even at temperatures greater than the magnetic ordering
temperature(T$_C$)\cite{tokura,salamon,betracco}. In Fig. 3(a) we
show the $\rho(T)$ in the temperature range of 10-300 K of a LSMO
film deposited on SrTiO$_{3}$. The temperature dependence seen
here is consistent with the canonical behavior stated in the
preceding line. Interestingly, the polycrystalline films deposited
at 800 $^{0}$C on YSZ also shows a similar behavior, although with
a higher overall resistivity, which can be attributed to grain
boundary scattering in a homogeneous polycrystalline film. The
temperature dependence of resistivity in metallic LSMO is given
by\cite{snyder}
\begin {eqnarray}
\rho (T) = \rho _1  + AT^2  + A_{4.5}T^{4.5},
\end {eqnarray}
where $\rho_1$ is the residual temperature-independent resistivity
due to scattering by impurities and defects. The second term in
Eq. (1) is due to electron-electron scattering, and third term
describes the second order electron-magnon
scattering\cite{snyder}. The parameters extracted by fitting Eq.
(1) to the data of Fig. 3(a) are listed in the Table 1 and they
match well with the values reported in
literature\cite{snyder,chen}. The temperature dependence of the
field-cooled (FC) magnetization M(T) measured at 100 Oe of the
film deposited at 800 $^{0}$C on YSZ is shown in the inset of Fig.
3(a). The magnetization near the Curie temperature goes to zero as
M(T) $\sim$ (T-T$_C$)$^{\beta}$, with $\beta$ $\approx$ 0.38
$\pm$0.01 and T$_C$ $\approx$ 350 K. The critical exponent $\beta$
compares well with the value ($\approx$ 0.37 $\pm$ 0.04) for
single crystal La$_{2/3}$Sr$_{1/3}$MnO$_{3}$ \cite{gosh}. In the
same inset we also plot the FC magnetization of a film deposited
at 700 $^{0}$C on YSZ. While the magnetic ordering temperature
T$_C$ of the 700 and 800 $^{0}$C deposited films is nearly the
same, a remarkably different temperature dependence of resistivity
emerges for the former, as shown in Fig. 3(b). The $\rho(T)$ is
now thermally activated down to T$_{P}$ and then a metallic
behavior develops in the temperature window of 170 - 50 K followed
by reentry into a semiconducting regime at still lower
temperatures. The resistivity of these films at T $\leq$ T$_{P}$
also drops significantly on application of a small magnetic field
with a concomitant shift of T$_{P}$ to higher temperatures as
shown in the inset of Fig. 3(b).

We now examine the $\rho$(T) data of Fig. 3(b) in the light of
structural and topographical information. Our AFM results clearly
indicate a two-component medium where well formed but randomly
oriented grains of LSMO are separated by an intergranular
material, which is presumed to be a dirty ferromagnetic metal with
a range of magnetic ordering temperatures lower than the T$_C$ of
bulk LSMO. A depression in T$_C$ of large bandwidth manganites
such as LSMO due to distortion in bond angles and bond lengths,
and oxygen deficiency is well established
experimentally\cite{park,shuller}. Such deviations from the ideal
structure are common at the surfaces of grains in transition metal
oxides\cite{gupta,zeise2}.  A pictorial view of our model is
presented in Fig. 4. We divide the resistivity curve into three
zones. In the Zone-1 (T$_{P}<T<T_{C}$), we have nanometer scale
($\approx$ 80 nm) feromagnetic granules of the Curie temperature
T$_{C}^{G}$ ($\approx$ 350 K) embedded in a paramagnetic
insulator. This is the intergranular material with a magnetic
ordering temperature T$^{IG}$$_{C}$ $\approx$ T$_{P}$, which is
consistent with the observation that the T$_C$ of LSMO can be
reduced significantly by oxygen deficiency and
disorder\cite{park,shuller}. Then comes Zone-2 of temperature
(T$_{min}<T\leq T_{P}$) where the intergranular disordered LSMO is
a dirty metal with ferromagnetic ordering. Finally in the Zone-3
($T\leq T_{min}$) Kondo-type scattering from the blocked Mn-spins
in the intergranular dirty ferromagnet leads to the divergence of
resistivity  on cooling below T$_{min}$.

We now discuss electron transport in each of these regimes in
detail and augment this intuitive model with more data. The
resistivity in Zone-1 is thermally activated but with a negligibly
small temperature dependent activation energy, suggesting phonon
mediated variable range tunneling through the paramagnetic
insulator. The insignificant field dependence of resistivity in
Zone-1 [see Fig. 3(b)] can be attributed to spin-flip scattering
of spin-polarized holes as they move between two neighboring
clusters via hopping through the localized states of the
paramagnetic insulator where each localizing state has a moment
(Mn$^{3+}$ polaron with S = 3/2). The spin-flip scattering does
not permit a spin-selective transport between two ferromagnetic
grains, which is essential for large magnetoresistance (MR). In
magnetically inactive insulating media such as SiO$_{2}$,
Al$_{2}$O$_{3}$, MgO etc. where spin-selectivity is not blocked,
one, however, sees a large MR. Typical examples of such system are
cermet film like Ni-SiO$_{2}$, Co-SiO$_{2}$ and Fe-SiO$_{2}$
etc.\cite{sheng,abeles,honda,milner}

We argue that in the vicinity of  T $\approx$ T$_{P}$, a
sufficient fraction of the  intergranular material undergoes
magnetic ordering to a ferromagnetic phase. Like other double
exchange manganites, the ordered phase is presumably metallic with
a large value of resistivity, which is consistent with observed
inverse scaling between  $\rho(0)$ and T$_C$ of the
manganites\cite{tokura,salamon}. The large magnetoresistance near
T$_{P}$ and the shift of T$_{P}$ to higher temperatures on
increasing the field are also consistent with this picture of a PM
to FM transition accompanied by a transition in electrical
conductivity. In the metallic regime (Zone-2 of Fig. 4(c)), the
system can be viewed as a microscopic mixture of two materials,
grain G and intergranular IG, with respective resistivity
$\rho_{G}$ and $\rho_{IG}$ . The temperature dependence is
dominated entirely by intrinsic transport in the more resistive
intergranular dirty metal. We believe that tunneling dominated
transport, as argued by some workers for polycrystalline films of
LSMO, is unlikely to occur in such cases because one does not see
a Coulomb blockade regime in current-voltage characteristics even
at fairly low-temperatures (5 K) which remain linear even at very
small bias voltages. It is worth pointing out that electron
transport in granular manganites has often been compared with the
data on thin films of insulating matrix system such as
Co-SiO$_{2}$, Ni-SiO$_{2}$ and Ni-SiO$_{2}$. The resistivity of
such systems at low-temperature varies
as\cite{milner,sheng,abeles,honda};
\begin {eqnarray}
\rho  = \rho _0 \exp \left( {\frac{{T_0 }} {T}} \right)^{1/2},
\end {eqnarray}
where the exponential factor T$_{0}$ depends on the electrostatic
charging energy of the metallic grains (Ni or Co), width of
tunneling barrier and the barrier height. These systems, however,
do not show the type of metallic behavior seen in Zone-2 of Fig.
4(c).

In Fig. 5(a) we plot the semiconductor-like resistivity of Zone-3
as a function of T$^{-1/2}$. For zero-field case the value of
T$_{0}$ is found to be $\approx$ 1.07 K (0.09 meV). However, a
calculation based on the charging energy E$_C$=e$^2$/2C, with
C=4$\pi\epsilon_0\epsilon$r, r$\approx$ 80 nm and $\epsilon$
$\approx$ 10 gives E$_C$ $\approx$ 0.9 meV, which is an order of
magnitude larger than T$_0$. Although T$_0$ also depends on
barrier width and height, such large mismatch in measured T$_{0}$
and calculated E$_{C}$ cannot be accounted for with admissible
variations in barrier height and its width. In the following, we
argue that the transport in Zone-3 is not due to ballistic
tunneling through a clean wide-band-gap insulator like SiO$_{2}$
or Al$_{2}O_{3}$ as in cermet films, but a consequence of
Kondo-type scattering in the intergranular spin-disordered dirty
metal.

To understand the temperature dependence of resistivity in more
detail one can consider a two-component system whose effective
resistivity is given by the effective medium
approximation\cite{kirkpitrick,ju2} as
\begin {eqnarray}
f\left( {\frac{{\rho _M  - \rho _{eff} }}{{2\rho _M  + \rho _{eff}
}}} \right) + (1 - f)\left( {\frac{{\rho _{DM}  - \rho _{eff}
}}{{2\rho _{DM}  + \rho _{eff} }}} \right) = 0.
\end {eqnarray}
Here $\emph{f}$ represents the volume fraction of clean
ferromagnetic metallic regions whose resistivity is $\rho_M$ [ the
grains ] and (1-$\emph{f}$) the dirty ferromagnetic metal of
intergrain space having resistivity $\rho_{DM}$. However we have
used a simpler approach where the clean and dirty metallic regions
are connected in series. The effective resistivity is written as;
\begin {eqnarray}
\rho _{eff}  = f\rho _{M}  + (1 - f)\rho _{DM}.
\end {eqnarray}
We assume that the semiconducting behavior in Zone-3 is either due
to electron-electron (e-e) interaction or due to Kondo-scattering
in the small (1-$\emph{f}$) fraction of the intergranular dirty
metal. In the e-e interaction picture, the resistivity is given
by\cite{bhagat}
\begin {eqnarray}
\rho _{DM}  = \rho _2 (1 - BT^{1/2} )
\end {eqnarray}

where $\rho _2 $ is the residual resistivity contributed by
temperature independent scattering processes and $\emph{B }$  a
constant with value in the range of 5 - 20 $\times 10^{ - 4} $
K$^{-1/2}$ for disordered alloys. Substituting Eqs. (5) and (1)
into Eq. (4), we find
\begin {eqnarray}
\rho _{eff}  = f\rho _1  + fAT^2  + fA_{4.5} T^{4.5}  + (1 -
f)\rho _2  + \rho _2 (f - 1)BT^{1/2}
\end {eqnarray}

A best fit of the data in Zone-3 to Eq. (6)  is displayed in Fig.
5(b). The constants A and $A_{4.5}$ for the fit were taken from
the data on epitaxial LSMO thin film as listed in Table 1.
Furthermore, the metallic fraction has been set at $\emph{f}$
$\approx$ 0.9 by analyzing the AFM images of the polycrystalline
film (ellipsoidal grains with a $\approx$ b $\approx$ 30 nm and c
$\approx$ 80 nm with a grain boundary thickness of $\sim$ 1 nm).
However, even with such reasonable choices of $\rho_1$,
$\emph{A}$, $\emph{A}_{4.5}$ and $\emph{f}$ , poor quality of fit
obtained suggests that the electron-electron interaction picture
may not be appropriate to explain the low-temperature behavior of
$\rho$(T).

In an alternative approach considering a Kondo-type behavior of
resistivity in the dirty ferromagnetic metal, we can
write\cite{hasegawa}
\begin {eqnarray}
\rho _{DM}  = \rho_3  + \delta T^2  + \gamma \ln T
\end {eqnarray}
Where $\rho_3 $ is the temperature independent residual
resistivity and  the T$^2$ term is due to electron-electron
scattering as observed in typical metals. The third term with
coefficient $\gamma$ corresponds to the Kondo-type scattering
process, which in the present case will be due to disordered
Mn-spins. From Eq. (7) we can write the temperature corresponding
to resistivity minimum as;
\begin {eqnarray}
T_m  = \left( { - \frac{\gamma }{{2\delta }}} \right)^{1/2}.
\end {eqnarray}

Substituting Eq. (7) and Eq. (1) into Eq. (4), and rearranging the
coefficients the effective resistivity can be written as;
\begin {eqnarray}
\rho _{eff}  = \alpha  + f\rho _1  + fAT^2  + fA_{4.5}T^{4.5}  +
(1 - f)\delta T^2  + (1 - f)\gamma \ln T,
\end {eqnarray}
where the constant $\alpha$ (= $(1-\emph{f})\rho_3$) depends on
the residual resistivity of the intergranular dirty metal. A
fitting of Eq. (9) to the experimental data in the temperature
range 50 to 5 K is shown in Fig. 5(c) with fixed $\rho_1$,
$\emph{f}$, $\emph{A}$ and $\emph{A}_{4.5}$ derived from $\rho$(T)
of epitaxial films and AFM images shown in Fig. 2(a). The fitting
parameters for different field in the temperature range 50 to 10 K
are listed in Table. 1. The resistivity minimum temperature
T$_{m}$ calculated using Eq. (8) is 52.2 K. The larger value of
parameter $\delta$ compared to that of $\emph{A}$ is due to  the
fact that $\emph{A}$ increases rapidly with oxygen deficiency and
strain. This become clear from the order of magnitude larger value
of A in film deposited on YSZ at 800 $^0$C compared to those on
STO of the same T$_{D}$.  The films on YSZ are polycrystalline and
likely to have oxygen deficiency at the grain boundaries.

To understand the effect of magnetic field, the temperature
dependent resistivity measured at different fields is also fitted
to Eq. (9). The best fit to the data taken at 0.3 T is shown in
Fig. 5(c). The coefficient $\gamma$ of the ln $\emph{T}$ term in
Eq. (7) calculated from the fitting is shown in Fig. 5(d). The
negative sign of the coefficient $\gamma$ is consistent with a
Kondo-type picture. Thus application of magnetic field suppresses
spin- disorder and hence spin-dependent-scattering and  leads to
negative magnetoresistance.

To understand the origin of magnetoresistance further,
particularly in a low field regime where moment of the grains are
not fully aligned, we look at the correlation between isothermal
magnetization and low-field magnetoresistance of these films in
the temperature regime corresponding to Zone-3. A typical set of
data taken at 20 K for the samples deposited at 700 and 800 $^{0}$
C acquired in a configuration where field and current are coplanar
but orthogonal to each other is shown in Fig. 6. The MR has been
defined as;

\begin {eqnarray}
MR(\% ) = \frac{{R(H) - R(0)}}{{R(0)}} \times 100,
\end {eqnarray}
where R(H) and R(0) are resistances in the presence of field and
in zero-field respectively. The MR is initially positive and
increases with field, peaking for a field of H$_{P}$ ($\approx$
222 Oe). On increasing the field beyond H$_{P}$, the MR becomes
negative and reaches a value $\approx$ 19 $\%$ for a small field
of 0.3 T. However, initial $\sim$ 70 $\%$ of the drop in
resistance occurs for field  H $\leq$ 0.15 T followed by a linear
increase at higher fields. The MR in these films were found out to
be symmetric for both positive and negative field direction. The
hysteresis in MR can also be understood considering the fact that
for lower fields the magnetization vector ($\textbf{M}$) of
individual grains are aligned along their easy axis and for field
higher than H$_C$ the grains get aligned along the field direction
resulting in a parallel low resistance state\cite{soumen}. In Fig.
6 we have also plotted the field dependence of magnetization to
emphasize the correlation between coercive field ( H$_C$ ) and the
field ($\pm$ H$_P$) at which MR shows peaks on field reversal. We
notice that while for the films with T$_D$= 800 $^0$C, H$_P$
$\approx$ H$_C$, where as in the film deposited at 700 $^0$C H$_P$
$>$ H$_C$. Similar inequality has also been seen in other
polycrystalline half-metallic perovskite oxides like
Sr$_{2}$FeMoO$_{6}$ (SFMO)\cite{sarma}. The distinctive feature in
the present case is that the intergrain material is a
spin-disordered dirty metal. However, the small mismatch (~100 Oe)
between H$_P$ and H$_C$ can also be understood by considering the
fact that in MR measurement the current circulates only a fraction
of the magnetic grains which fall on the percolating path where as
the magnetization is a bulk measurement which reflects the average
response of the whole sample\cite{bose}. The 800 $^{0}$C deposited
sample showed a better H$_C$ vs H$_P$ matching may be due to
better connectivity between grains therefore a larger fraction of
the sample contributes to the MR measurement.

\subsection{Anisotropic magnetoresistance (AMR)}
Although the anisotropy in magnetoresistance seen in ferromagnetic
metals may have different origin  involving  increased pathlength
due to the Lorentz force, Fermi surface effects etc., the
anisotropic scattering due to spin-orbit interaction has been
argued to be the main contributor to
AMR\cite{campbell,mcguire,malozemoff}. The anisotropic
magnetoresistance in polycrystalline ferromagnets typically
follows a $\cos ^2 \theta$ dependence, where $\theta$ is the angle
between the current and magnetization vector of the sample.

This cos$^2$$ \theta $ type AMR in transition metal ferromagnets
and alloys is understood theoretically through scattering of
s-electrons in to empty d-states through spin-orbit interaction,
H$_{SO}$ $\approx$ $\triangle_{SO}LS$, where L and S are orbital
and spin angular momentum respectively and $\triangle_{SO}$ is the
spin-orbit coupling parameter. In double exchange manganites the
carrier lie in a band formed by itinerant e$_g$-band which is
strongly coupled to localized t$_{2g}$ band via strong Hund's
coupling. Due to the itinerant nature of the e$_g$ band the
orbital moment (L = 0) is strongly quenched. However a small
spin-orbit interaction of conduction carriers arises from the
hybridized O 2p orbital. Moreover, in polycrystalline thin film of
manganite the oxygen vacancy and deformations from ideal crystal
structure at grain boundaries might result in an enhanced
AMR\cite{mathieu}. The AMR at 3 kOe for two LSMO films of
different texture is shown in Fig. 7. In case of the
polycrystalline film deposited at 700 $^{0}$C, the AMR can be
fitted to an equation of the type AMR ($\%$) = A + B cos$^2$($
\theta $) with A=2.94 $\pm$ 0.03 and B=-2.98 $\pm$ 0.05. The
cos$^2$$ \theta $ dependence reflects the spin-orbit coupling type
AMR behavior. A more complicated functional behavior was found for
the film deposited at 800 $^{0}$C. Additional peaks were also
found at $\theta$ = 60 , 120 and 180 degrees which can be
attributed to pinning of the magnetization vector along certain
direction in the plane of the film. Interestingly, the AMR of
these films is order of magnitude smaller than that of the
polycrystalline films deposited at 700 $^{0}$C. The larger AMR in
case of polycrystallline samples may be understood in terms of
stress fields at the grain boundaries. The strain at the grain
boundaries may change intrinsic properties such as crystal-field
splitting locally thereby affecting AMR amplitude drastically.

\section{Conclusions}

We have presented a detailed study of magneto-transport in well
characterized polycrystalline films of LSMO having different
degrees of granularity and crystalline texture engineered by
controlling the growth temperature. The granularity in films
deposited at 700 and 800 $^{0}$C was established through AFM
imaging and $\theta-2\theta$ X-ray diffraction. Although both the
films showed ferromagnetic ordering at T$_C$ $\approx$ 350 K, the
temperature dependence of the resistivity ($\rho$(T)) of these
films was dramatically different. An order of magnitude higher
resistivity, lower metal-insulator transition temperature and a
low-temperature upturn of $\rho(T)$ was found in 700 $^{0}$C
deposited film because of the enhanced grain boundary scattering.
A model for magnetotransport has been proposed by considering the
film equivalent to  a composite system of metallic LSMO granules
embedded in a matrix of disordered metal formed from  oxygen
deficient and strained LSMO at the grain boundaries. The
temperature dependence of resistivity of this two-component
composite is governed by the more resistive disordered metal which
undergoes the double-exchange driven magnetic ordering and
metallic transition at T$_P$ $\approx$ 170 K . At temperature
below 50 K, an upturn in the resistivity is seen presumably due to
a Kondo-type scattering from the Mn-spins at the grain boundaries.
A large magnetoresistance ( $\approx$ 19 $\%$) for a field of 0.3
T is seen at 10 K in films of T$_D$ = 700 $^{0}$C in comparison to
only 0.4 $\%$ MR for 800 $^{0}$C deposited films. The former also
shows a two orders of magnitude higher anisotropic
magnetoresistance. The dramatically different magnetotransport in
such films presumably results from their two-component nature
where spin-dependent scattering in the intergranular regions is
dominant.

\begin{acknowledgements}
This research has been supported by grants from the Ministry of
Communication and Information Technology (MCIT) and Board of
Research in Nuclear Sciences (BRNS), Government of India. P. K.
Muduli acknowledges financial support from the Council for
Scientific and Industrial Research (CSIR), Government of India.
\end{acknowledgements}

\clearpage
\begin{table}[h]
\caption{Fitting results of Eq. (1) and Eq. (9)} \label{tab1}
\begin{tabular}{llllllllllllllllll}
\toprule
Film &&$\rho_1$ ($\Omega$-cm)&&A ($\Omega$-cm K$^{-2}$) &&&&&$A_{4.5}$ ($\Omega$-cm K$^{-4.5}$)\\
\colrule
SrTiO$_{3}$ &&1.1$\times$10$^{-4}$&&7.9$\times$10$^{-9}$&&&&&4.3$\times$10$^{-15}$\\
YSZ &&2.2$\times$10$^{-4}$&&1.5$\times$10$^{-8}$&&&&&6.2$\times$10$^{-15}$\\
\colrule
Field (T)&&$\alpha$ ($\Omega$-cm)&&$\delta$($\Omega$-cm K$^{-2}$) &&&&&$\gamma$ ($\Omega$-cm)\\
\colrule
0 &&8.3$\times$10$^{-1}$&&2$\times$10$^{-5}$&&&&&-0.109\\
0.1 &&7.1$\times$10$^{-1}$&&2$\times$10$^{-5}$&&&&&-0.089\\
0.2 &&6.6$\times$10$^{-1}$&&2$\times$10$^{-5}$&&&&&-0.083\\
0.3 &&6.4$\times$10$^{-1}$&&2$\times$10$^{-5}$&&&&&-0.078\\

\botrule
\end{tabular}

\end{table}

\clearpage
\begin{figure}[h]
\begin{center}
\includegraphics [width=12cm]{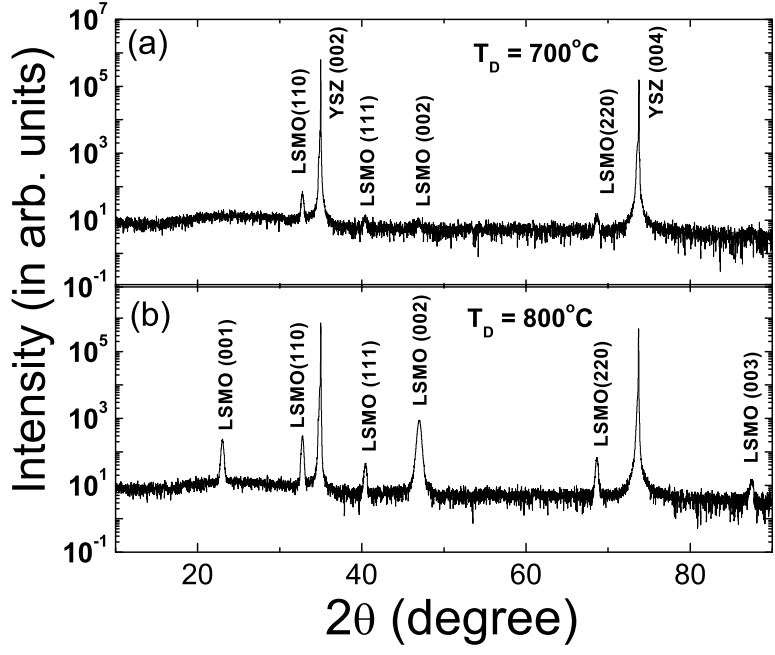}%
\end{center}
\caption{\label{fig1} (a,b) $\theta$ - 2$\theta$  X-ray
diffraction scans of 100 nm thick LSMO films deposited on (100)
YSZ substrate at deposition temperature T$_{D}$ $=$ 700 $^{0}$C
(a) and  T$_{D}$ $=$ 800 $^{0}$C (b). The presence of  (00l),
(hk0) and (hkl) reflections confirm a truly polycrystalline
growth. All the orientations of LSMO thin film in this paper have
been described in terms of pseudocubic unit cell.}
\end{figure}
\clearpage
\begin{figure}[h]
\begin{center}
\includegraphics [width=12cm]{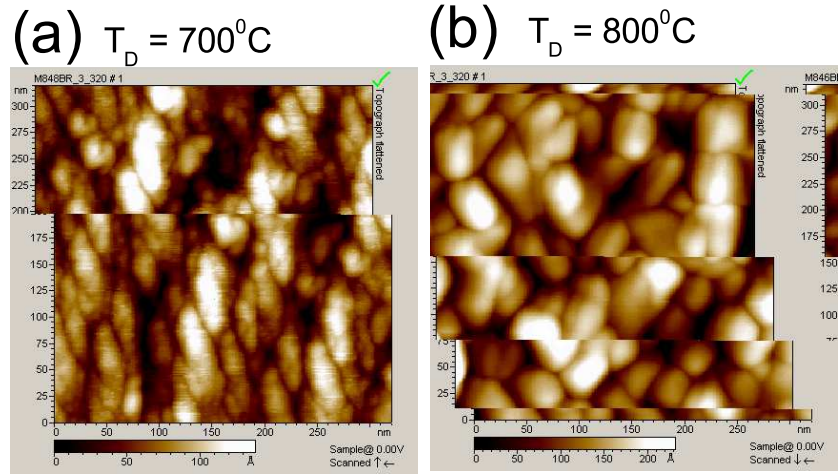}%
\end{center}
\caption{\label{fig2}(a,b) AFM photographs of 100 nm thick LSMO
films deposited on (100) YSZ substrate at 700 and 800 $^{0}$C are
shown in panel 'a' and 'b' respectively. The images were taken at
room temperature in non-contact mode. }
\end{figure}

\clearpage
\begin{figure}[h]
\begin{center}
\includegraphics [width=12cm]{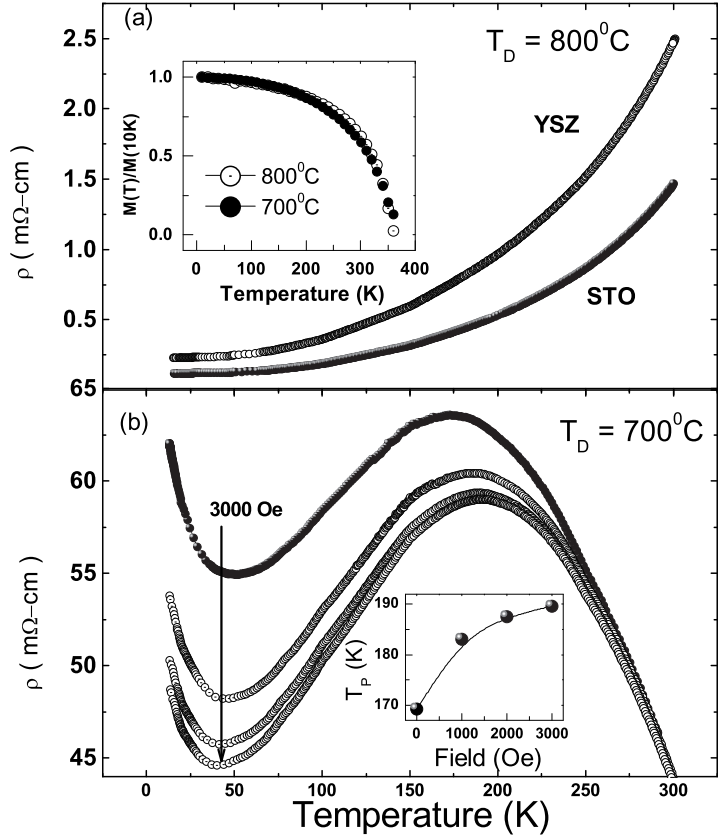}%
\end{center}
\caption{\label{fig3} (a) Temperature dependence of resistivity of
a 100 nm thick LSMO film  deposited at 800 $^{0}$C on (100) YSZ
and (100) STO substrates is shown in panel 'a'. The temperature
dependence of field-cooled magnetization of the film on YSZ
deposited at 700 and 800$^{0}$C is shown by closed and open
circles respectively in the inset. Panel 'b' shows temperature and
field dependence of resistance of a 100 nm thick LSMO film
deposited at 700 $^{0}$C on (100)YSZ. The inset shows the field-
dependence of metal insulator-transition temperature T$_{P}$.}
\end{figure}
\clearpage
\begin{figure}[h]
\begin{center}
\includegraphics [width=12cm]{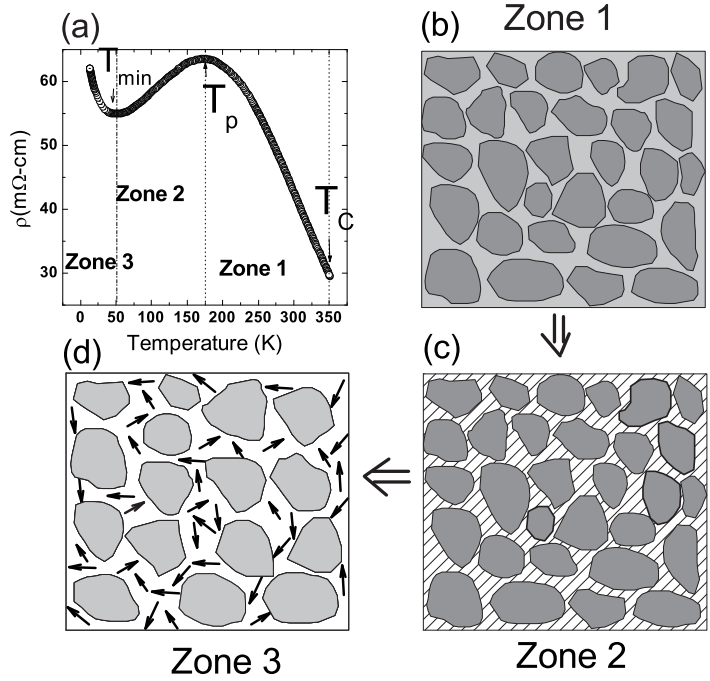}%
\end{center}
\caption{\label{fig4} Panel 'a' shows the zero-field resistivity
of the 700 $^{0}$C deposited polycrystalline film in the
temperature range of 10 to 350 K. The curve has been divided into
three temperature zones. Panel 'b', 'c' and 'd' are sketches of
granular structure corresponding to Zone 1 to 3 respectively.
 }
\end{figure}

\clearpage
\begin{figure}[h]
\begin{center}
\includegraphics [width=12cm]{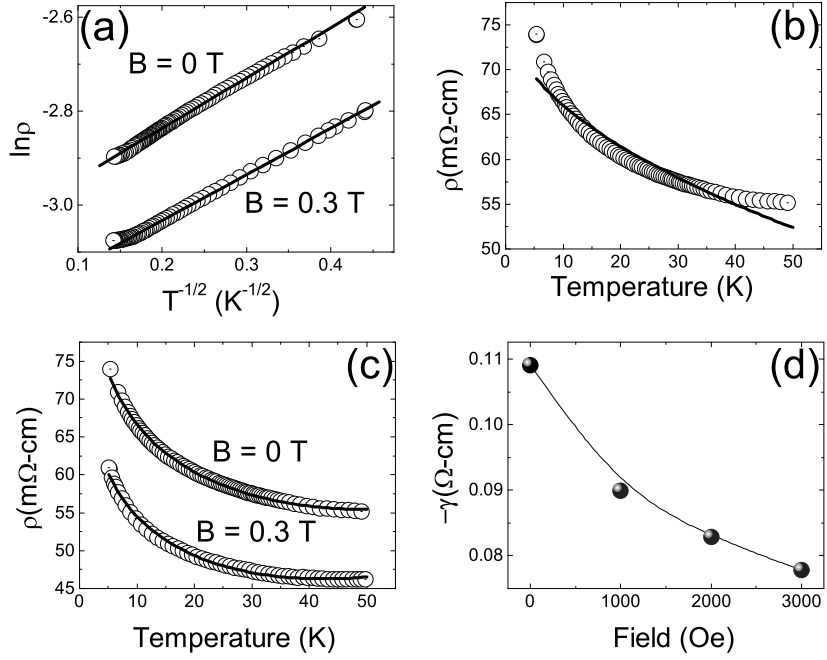}%
\end{center}
\caption{\label{fig5 } Panel 'a' shows the exp [T$^{-1/2}$]
depedence of the resistivity of a 100 nm thick polycrystalline
film at T $\leq$ 50 K for zero and 0.3 T field measurement. The
solid line is a fit to the expression $ \rho (T)\sim \exp \left(
{\frac{{T_0 }}{T}} \right)^{1/2}$. Panel 'b' and 'c' show the same
data fitted to Eq. (5) and (7) of the text. The field dependence
of coefficient $\gamma$ of Kondo-type scattering found by fitting
Eq. (7) to the data in Fig. 3(b) is shown in panel 'd'. }
\end{figure}

\clearpage
\begin{figure}[h]
\begin{center}
\includegraphics [width=12cm]{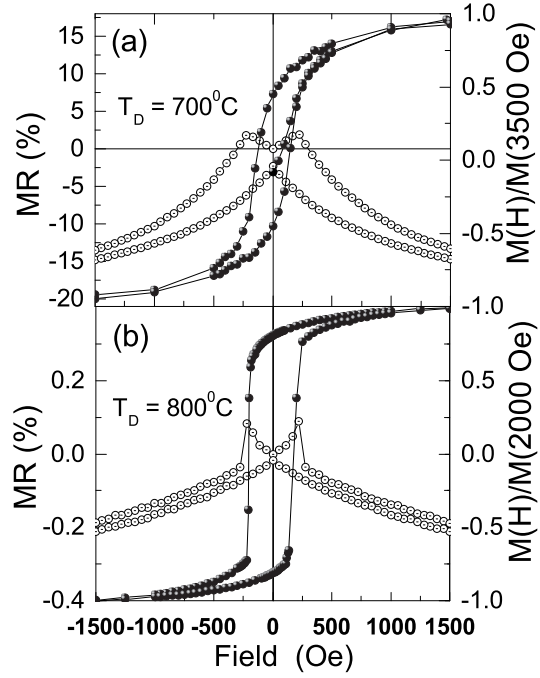}%
\end{center}
\caption{\label{fig6} Field dependence of magnetoresistance [$ MR
= \frac{{R(H) - R(0)}}{{R(0)}} \times 100 $ ] at 20 K for LSMO
films deposited at 700 and 800 $^{0}$ C is shown in panel 'a' and
'b' respectively. The field and current in these measurements are
coplanar but orthogonal to each other. The M(H) loops of the same
film measured at 20 K in coplanar field are also shown. }
\end{figure}
\clearpage
\begin{figure}[h]
\begin{center}
\includegraphics [width=12cm]{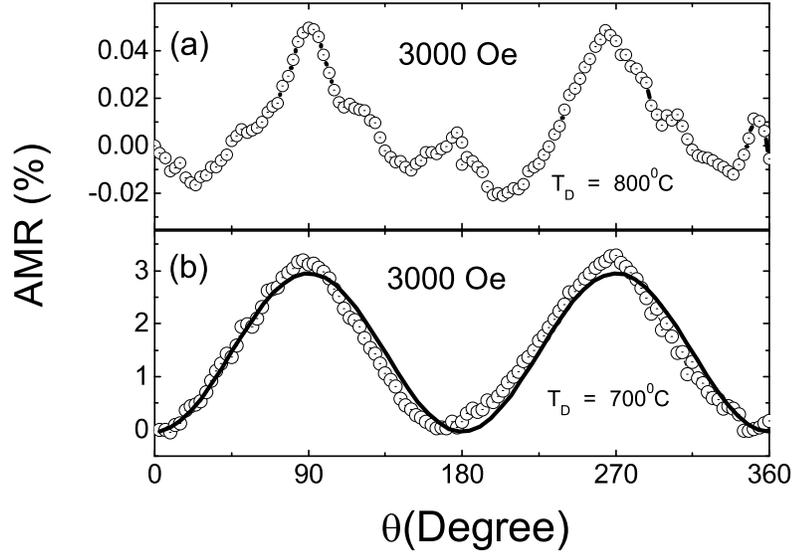}%
\end{center}
\caption{\label{fig7} Angular dependence of magnetoesistance (AMR)
of two LSMO films measured at 50 K. Panels 'a' and 'b' are for the
film deposited at 800 and 700 $^{0}$C respectively on YSZ(100)
substrate. The angle $\theta$ is the angle between the applied
magnetic field and the direction of current flowing through the
sample. The field and current are coplanar. Solid line in the
panel 'b' is a fit of the type AMR($\%$) $\approx$
cos$^{2}$$\theta$.}
\end{figure}

\end{document}